\documentclass[twocolumn]{elsart}
\usepackage{graphicx}
\usepackage{amsmath}
\usepackage{url}

\newenvironment{myitemize}%
  {\begin{list}{}
      {%
       \setlength{\leftmargin}{2em}
       \setlength{\rightmargin}{2em}
       \setlength{\listparindent}{0pt}
       \setlength{\itemsep}{0pc}
       \setlength{\parsep}{0pc}
       \setlength{\itemindent}{-1em}
       \setlength{\labelwidth}{0em}}}
  {\end{list}}

\begin{document}

\begin{frontmatter}
\title{Observation of the Ankle and Evidence for a High-Energy Break
    in the Cosmic Ray Spectrum}

\author[Utah]{R.U.~Abbasi}
\author[Utah]{T.~Abu-Zayyad}
\author[LANL]{J.F.~Amman}
\author[Utah]{G.~Archbold}
\author[Utah]{R.~Atkins}
\author[Adelaide]{J.A.~Bellido}
\author[Utah]{K.~Belov}
\author[Montana]{J.W.~Belz}
\author[Columbia]{S.Y.~Ben~Zvi}
\author[Rutgers]{D.R.~Bergman\thanksref{email}}
\author[Utah]{G.W.~Burt}
\author[Utah]{Z.~Cao}
\author[Adelaide]{R.W.~Clay}
\author[Columbia]{B.C.~Connolly}
\author[Utah]{W.~Deng}
\author[Adelaide]{B.R.~Dawson}
\author[Utah]{Y.~Fedorova}
\author[Utah]{J~Findlay}
\author[Columbia]{C.B.~Finley}
\author[Utah]{W.F.~Hanlon}
\author[LANL]{C.M.~Hoffman}
\author[Rutgers]{G.A.~Hughes}
\author[LANL]{M.H.~Holzscheiter}
\author[Utah]{P.~H\"{u}ntemeyer}
\author[Utah]{C.C.H.~Jui}
\author[Utah]{K.~Kim}
\author[Montana]{M.A.~Kirn}
\author[Utah]{E.C.~Loh}
\author[Utah]{M.M.~Maestas}
\author[Tokyo]{N.~Manago}
\author[LANL]{L.J.~Marek}
\author[Utah]{K.~Martens}
\author[NewMexico]{J.A.J.~Matthews}
\author[Utah]{J.N.~Matthews}
\author[Columbia]{A.~O'Neill}
\author[LANL]{C.A.~Painter}
\author[Rutgers]{L.~Perera}
\author[Utah]{K.~Reil}
\author[Utah]{R.~Riehle}
\author[NewMexico]{M.~Roberts}
\author[Tokyo]{M.~Sasaki}
\author[Rutgers]{S.R.~Schnetzer}
\author[Adelaide]{K.M.~Simpson}
\author[LANL]{G.~Sinnis}
\author[Utah]{J.D.~Smith}
\author[Utah]{R.~Snow}
\author[Utah]{P.~Sokolsky}
\author[Columbia]{C.~Song}
\author[Utah]{R.W.~Springer}
\author[Utah]{B.T.~Stokes}
\author[Utah]{J.R.~Thomas}
\author[Utah]{S.B.~Thomas}
\author[Rutgers]{G.B.~Thomson}
\author[LANL]{D.~Tupa}
\author[Columbia]{S.~Westerhoff}
\author[Utah]{L.R.~Wiencke}
\author[Rutgers]{A.~Zech}
                                               
\address[Utah]{University of Utah, Department of Physics and High
Energy Astrophysics Institute, Salt Lake City, Utah, USA}
\address[LANL]{Los Alamos National Laboratory, Los Alamos, NM, USA}
\address[Adelaide]{University of Adelaide, Department of Physics,
Adelaide, South Australia, Australia}
\address[Montana]{University of Montana, Department of Physics and
Astronomy, Missoula, Montana, USA}
\address[Columbia]{Columbia University, Department of Physics and
Nevis Laboratory, New York, New York, USA}
\address[Rutgers]{Rutgers - The State University of New Jersey,
Department of Physics and Astronomy, Piscataway, New Jersey, USA}
\address[NewMexico]{University of New Mexico, Department of Physics
and Astronomy, Albuquerque, New Mexico, USA}
\address[Tokyo]{University of Tokyo, Institute for Cosmic Ray
Research, Kashiwa, Japan}
\collaboration{The High Resolution Fly's Eye Collaboration}

\thanks[email]{To whom correspondence should be addressed.  E-mail:
\url{bergman@physics.rutgers.edu}}

\date{\today}

\begin{abstract}
  We have measured the cosmic ray spectrum at energies above
  $10^{17}$~eV using the two air fluorescence detectors of the High
  Resolution Fly's Eye experiment operating in monocular mode.  We
  describe the detector, PMT and atmospheric calibrations, and the
  analysis techniques for the two detectors.  We fit the spectrum to
  models describing galactic and extragalactic sources.  Our measured
  spectrum gives an observation of a feature known as the ``ankle''
  near $3 \times 10^{18}$ eV, and strong evidence for a suppression
  near $6\times{10}^{19}$~eV.
\end{abstract}
\end{frontmatter}

\section{Introduction}

The highest energy cosmic rays yet detected, of energies up to and
above $10^{20}$ eV, are interesting in that they shed light on two
important questions: how are cosmic rays accelerated in astrophysical
sources, and how do they propagate to us through the cosmic microwave
background radiation (CMBR)\cite{cmbr}?  The acceleration of cosmic
rays to ultra high energies is thought to occur in extensive regions
of high magnetic fields, regions which are expanding at relativistic
velocities\cite{acceleration}.  Such regions are rare and are to be
counted among the most violent and interesting objects in the
universe.

Once accelerated, interactions between the ultra high energy cosmic
rays (UHECR) and the CMBR cause the cosmic rays to lose energy.  The
strongest energy loss mechanism comes from the production of pions in
these CMBR interactions at UHECR energies above about $6\times
10^{19}$ eV.  This energy loss mechanism produces the
Greisen-Zatsepin-K'uzmin (GZK) suppression\cite{gr,zk}.  In addition,
$e^+ e^-$ production in these same interactions provides a somewhat
weaker energy loss mechanism above a threshold of about $5\times
10^{17}$ eV.  A third important energy-loss mechanism at all energies
comes from universal expansion.

In previous publications\cite{prl,app}, we have reported on our
measurements of the cosmic ray spectrum using data collected
independently, in monocular mode, by the two detectors of the High
Resolution Fly's Eye experiment (HiRes).  We here report on an updated
measurement of the flux of UHECR, covering an energy range from
$2.5\times 10^{17}$ eV to over $10^{20}$ eV, using a significantly
larger data set for the HiRes-II detector.  With the improved
statistical power available in this data, we study two features in
this spectrum: a break in the spectral slope at $3 \times 10^{18}$ eV,
called the ``ankle''\cite{fester,agasa,havpark}, and a steepening of
the spectrum near the threshold for pion production.

The HiRes experiment performs a calorimetric measurement of the energy
of cosmic rays.  UHECR produce extensive air showers (EAS) when they
enter the atmosphere.  The HiRes detector collects the fluorescence
light emitted by EAS as they propagate through the atmosphere.
Charged particles in the shower excite nitrogen molecules which
fluoresce in the ultraviolet (300 to 400 nm).  The fluorescence yield
is about five photons per minimum ionizing particle per meter of path
length\cite{fl_yield}.  As an EAS propagates through the atmosphere,
the detector measures the number of photons seen as a function of time
and angle.  From this information, we reconstruct the geometry of the
shower and the solid angle subtended by the detector from each point
of the shower.  From the number of photons collected, we reconstruct
the number of charged particles in the shower as a function of the
depth of the atmosphere traversed.  We integrate the energy deposited
in the atmosphere\cite{csong} to find the energy of the primary cosmic
ray.

UHECRs are thought to be protons or heavier nuclei up to iron.  While
nucleus-nucleus collisions are complex, the general features of the
interaction can be understood in terms of a simple superposition
model.  In this model each nucleon generates an independent EAS.  The
superposition of many, lower energy showers will result in an EAS with
different statistical properties than an EAS produced by one high
energy proton.  This allows one to measure the composition of the
primary cosmic rays on a statistical basis.

\section{The HiRes Detectors}

The HiRes detectors have been described extensively
elsewhere\cite{hr1_det,hr2_det}.  In brief, they consist of spherical
mirrors, of area 5.1 m$^2$, which collect the fluorescence light and
focus it onto a cluster of 256 photomultiplier tubes arranged in a 16
x 16 array.  Each tube in the cluster views about one square degree of
the sky.  Time and pulse height information are collected from each
tube.  The HiRes detectors trigger on and reconstruct showers that
occur within a radius of about 35 km.

The HiRes-I detector is located atop Little Granite Mountain on the
U.S. Army Dugway Proving Ground in west-central Utah.  It consists of
21 mirrors, and their associated phototube arrays, arranged in one
ring, observing from 3 degrees to 17 degrees in elevation and
providing almost complete coverage in azimuthal angle.  The detector
uses a sample-and-hold readout system which integrates phototube
pulses for 5.6 $\mu s$.  This is long enough to collect the signal
from all cosmic ray showers of interest.

The HiRes-II detector is located on Camel's Back Ridge, also on Dugway
Proving Ground, about 12.6 km SW of HiRes-I.  It consists of 42
mirrors, arranged in two rings, covering from 3 to 31 degrees in
elevation and almost the whole azimuthal angle range.  This detector
uses a flash ADC (FADC) readout system with a 100 ns sampling time.

In this article, we present data collected from June 1997 to February
2003 for HiRes-I, and from December 1999 through September 2001 for
HiRes-II.  For HiRes-II, this is about four times the data that was
reported on previously.  We collect data on nights when the moon is
down for three hours or more.  In a typical year each detector
collects up to about 1000 hours of data.

The weather is clear about $2/3$ of the time at the HiRes sites.
Since clouds can reduce the experiment's aperture, we record the
existence of clouds by operator observations, infrared cameras, and
evidence from data collected by the detector (this consists primarily
of the upward going laser and flasher pulses, used to measure the
atmospheric conditions, which have a distinct signature upon
encountering a cloud; actual cosmic rays also appear emerging from
clouds).  Only data from those nights in which the aperture is not
reduced by cloud cover are used in our spectrum measurements.

\section{Calibration}

The two most important calibrations we perform are of the
photomultiplier tube (PMT) gains\cite{yag,hr_cal}, and of the clarity
of the atmosphere\cite{hr_atm}.  We use a stable xenon flash lamp,
carried to each detector and used to illuminate the photomultiplier
array, to find PMT gains.  The xenon lamp produces a light intensity
of about 10 photons per mm$^2$ at the face of the PMTs; this intensity
is traceable to NIST-calibrated photodiodes and is stable to about 2\%,
flash-to-flash.  Separate calibrations of PMT gains using
photoelectron statistics and using the absolute light intensity of the
xenon flash lamp agree within uncertainties.  Xenon flash lamp data
are collected about once a month.  A second calibration system, using a
frequency-tripled YAG laser, is used to monitor phototube gains on a
night-to-night basis.  We estimate that the relative calibration
techniques are accurate to about 3\% with an absolute calibration
uncertainty of about $\pm 10$\%.

The atmosphere is our calorimeter, but it is also the medium through
which fluorescence light propagates to the detectors.  To calculate
the number of fluorescence photons emitted by a cosmic ray shower, we
must understand the way in which the atmosphere scatters this light
between the EAS and the detector.  The molecular component of the
atmosphere is quite constant, with only small seasonal variations, and
the Rayleigh scattering it produces is well understood.  The aerosol
content of the atmosphere can vary considerably over time, and with
it, the amount of light scattered and its angular distribution.

To measure these quantities, we perform an atmospheric calibration
using YAG lasers operating at wavelength $\lambda=355$ nm.  At each of
our two sites, we have a steerable beam laser which is fired in a
pattern of shots that covers the detector's aperture, and which is
repeated every hour.  The scattered light from the laser at one site
is collected by the detector at the other site.  The amount of
detected light is then analyzed to determine the scattering properties
of the atmosphere.  The properties that we measure are the vertical
aerosol optical depth (VAOD), the horizontal aerosol extinction
length, and the aerosol scattering phase function (the angular
distribution of the differential scattering cross section).

Because about half of the data from HiRes-I were collected before the
lasers were installed, we use average values of the measured
parameters in this analysis: a horizontal aerosol extinction length of
25 km (the average horizontal molecular extinction length is 17 km),
an average phase function, and a VAOD of
0.04\cite{prl,hr_atm,asz_thesis}.  The atmosphere at our sites is
quite clear: the average atmospheric correction to an event's energy
is about 10\% (see below for the effect on flux measurements).  We are
most sensitive to the value of VAOD.  The RMS of the VAOD distribution
is 0.02, and we use this RMS value as a conservative estimate of the
systematic uncertainty in this parameter.

\section{HiRes-II Data Analysis}

The analysis of the HiRes-II monocular data has been described
previously\cite{app}.  The data presented here were collected during
540 hours of good weather running, and consists of 21 million
triggers, mostly of random sky noise and events generated by
atmospheric lasers and other man-made light sources.  Events were
selected that satisfied the following criteria:
\begin{myitemize}
\item Angular speed $\le 11^\circ\ \mu s^{-1}$
\item Selected tubes $\ge 6$
\item Photoelectrons/degree $\ge$ 25
\item Track length $\ge 7^\circ$, or $\ge 10^\circ$ for events extending
above $17^\circ$ elevation
\item Zenith angle $\le 80^\circ$
\item In-plane angle $\le 130^\circ$
\item In-plane angle uncertainty $\le 30^\circ$
\item 150 $\le X_{\text{max}} \le 1200 \text{ g/cm}^2$, and is within
  50 g/cm$^2$ of begin visible in detector
\item Average \v{C}erenkov Correction $\le$ 70\%
\item Geometry fit $\chi^2$/d.o.f. $\le$ 10
\item Profile fit $\chi^2$/d.o.f. $\le$ 10
\item Minimal trigger from signal tubes required after March 2001
\end{myitemize}
These cuts remove events in which the monocular geometric
reconstruction is poor or in which the longitudinal profile cannot be
determined accurately.  The final event sample consisted of 2685
events covering an energy range from $1.6\times10^{17}$ eV
($\log_{10}E = 17.2)$ to $10^{20}$ eV.

The geometry of each event is reconstructed using the time and angle
information from the hit PMTs.  First a pattern recognition step is
performed to choose phototubes that lie on a line both in angle and in
time.  Next the plane that contains both the shower and the detector
is determined from the azimuth and elevation of hit tubes; the angle
of the shower in this plane is determined from a fit to phototube time
and angle information.  The resolution of shower-detector plane
determination is about $0.6^\circ$, and the in-plane angle uncertainty
is $5^\circ$ on average.

With the geometry determined, the profile of the number of charged
particles in the shower is calculated from the phototube pulse
heights.  Corrections are made for atmospheric scattering of the
light, and for other effects such as mirror reflectivity, phototube
quantum efficiency, etc.  A correction is made for the \v{C}erenkov
light produced by charged particles in the shower.  Both direct and
scattered \v{C}erenkov light contributions to the light seen by the
PMTs are calculated and subtracted.  The number of charged particles
is calculated from the fluorescence light at the shower using the
fluorescence yield and its pressure and temperature variation as given
by Kakimoto {\it et al}\cite{fl_yield}. The resulting shower
development profile, expressed as a function of slant depth, is fit to
the Gaisser-Hillas parameterization\cite{gh} (this has been seen to
fit UHE cosmic ray showers quite well\cite{csong,hr_prof}).  We
integrate over the Gaisser-Hillas function and multiply by the average
energy loss rate of 2.19 MeV/g/cm$^2$ to calculate the energy of the
primary cosmic ray.  We then correct for unobserved energy, mostly
neutrinos and muons which hit the ground.  This
correction\cite{asz_thesis}, which varies from 10\% at
$3\times10^{17}$ eV to 5\% at $10^{19}$ eV, is determined during the
Monte Carlo calculation of the aperture.  It is similar to the
calculation in reference \cite{csong}.

\begin{figure}
  \includegraphics[width=\columnwidth]{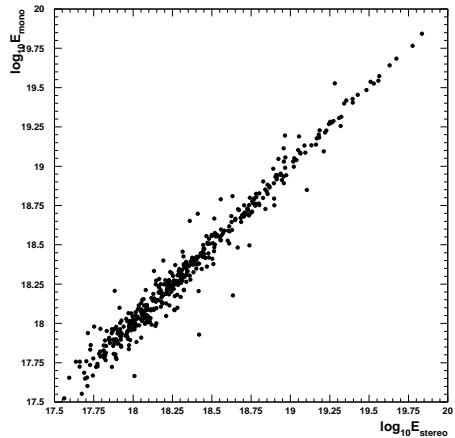}
  \caption{A scatter plot of the HiRes-II energy calculated
    using monocular geometry versus the energy calculated using the
    stereo geometry for those events observed in stereo.}
  \label{fig:hr2-stereo-v-mono}
\end{figure}

A fraction of the HiRes-II events are also observed by HiRes-I.  In
this case we perform a cross-check on our monocular determination of
the shower geometry.  Figure~\ref{fig:hr2-stereo-v-mono} shows a
scatter plot of the energy using monocular geometry versus the energy
using stereo geometry, for those events in which such a comparison is
possible.

The energy resolution, including statistical and systematic effects
and the uncertainty in reconstructing the shower geometry, has been
calculated in the Monte Carlo simulation\cite{asz_thesis}.  The
overall resolution is $\pm17\%$.  It improves from $\pm18\%$ below
$10^{18}$ eV to $\pm12\%$ above $10^{19}$ eV.
 
\section{Monte Carlo Simulation}

To calculate the aperture as a function of cosmic ray energy, a very
accurate Monte Carlo (MC) simulation of the experiment was
performed\cite{app}.  Two libraries of cosmic ray showers, one for
proton primaries and one for iron primaries, were generated using the
Corsika 5.61\cite{corsika} EAS simulation program and the QGSJet
01\cite{qgsjet} hadronic event generator.  Events from these libraries
were placed by a detector simulation in the vicinity of the HiRes-II
detector.  This program also simulated the fluorescence and
\v{C}erenkov light generated by the showers, and calculated how much
of this light would have been collected by the detectors.  A complete
simulation of the optical path, trigger, and readout electronics was
performed.  This simulation followed the experimental conditions that
pertained over the data-collection period.  The results were written
out in the same format as the data and analyzed by the same data
analysis program described above.  The stereoscopic energy spectrum of
the Fly's Eye experiment\cite{fester}, in the form of a broken power
law fit, and the composition measurements made by the HiRes/MIA hybrid
experiment\cite{hiresmia} and by HiRes in stereo\cite{gregpaper} were
used as inputs.

\begin{figure}
  \includegraphics[width=\columnwidth]{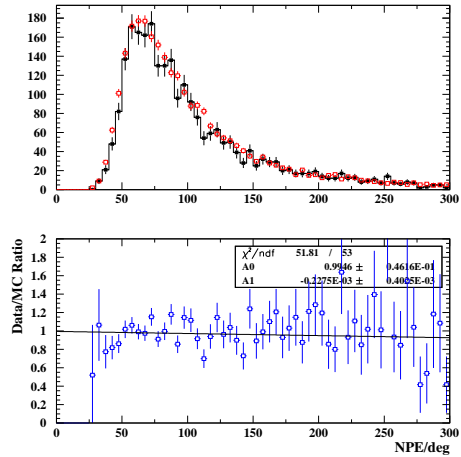}
  \caption{Comparison of HiRes-II data and MC for the photoelectrons
    per degree of track.  In the upper frame, the filled squares with
    the histogram are the data, the open squares are the MC.  The
    lower frame shows the ratio of data to MC for each bin.}
  \label{fig:npe}
\end{figure}

\begin{figure}
  \includegraphics[width=\columnwidth]{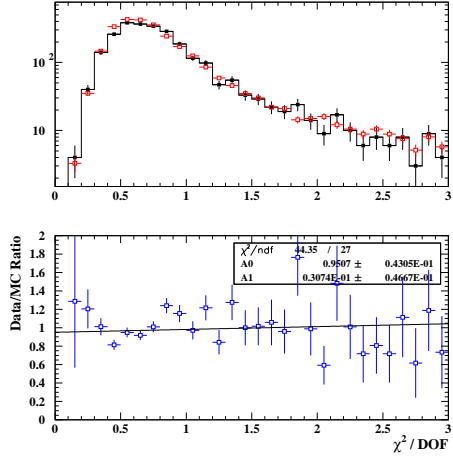}
  \caption{Comparison of HiRes-II data and MC for the $\chi^2$ of a
    fit to the time vs angle plot assuming a vertical shower.  In the
    upper frame, the filled squares with the histogram are the data,
    the open squares are the MC.  The lower frame shows the ratio of
    data to MC for each bin.}
  \label{fig:chisq}
\end{figure}

\begin{figure}
  \includegraphics[width=\columnwidth]{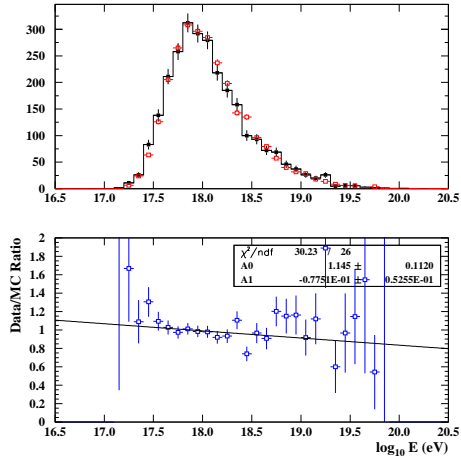}
  \caption{Comparison of HiRes-II data and MC for the reconstructed
    energy.  In the upper frame, the filled squares with the histogram
    are the data, the open squares are the MC.  The lower frame shows
    the ratio of data to MC for each bin.}
  \label{fig:energy}
\end{figure}

To convince ourselves that the MC simulation is accurate, we compare
many MC distributions of geometrical and kinematic variables to the
data.  The agreement in these comparisons is excellent and indicates
that we understand our detector.  Figure~\ref{fig:npe} shows the
brightness of showers: the number of photoelectrons per degree of
track.  The agreement between the data and MC simulation indicates
that the same amount of light is collected in the MC as in the data.
Figure~\ref{fig:chisq} shows the $\chi^2$ of a fit to the time vs.
angle plot from which we determine shower geometry.  The agreement
here indicates that the resolution of the MC is the same as that of
the data.  Figure~\ref{fig:energy} shows a histogram of the number of
events vs the logarithm of their energy in EeV.  The agreement here
shows that, when we use previous measurements of the spectrum and
composition in the MC, and a complete simulation of the acceptance, we
reproduce the experiment's energy dependence.

\section{HiRes-I Analysis}

The analysis of the HiRes-I monocular data has also been described
previously \cite{prl,app}.  The main difference from the HiRes-II
analysis is that, with only one ring of mirrors, most tracks are too
short to reliably determine the geometry from timing alone.  Although
the determination of the shower-detector plane is still excellent,
correlations between the fit distance to the shower and the fit
in-plane angle become large for short tracks.
 
A reconstruction procedure using the pulse height information in
addition to the tube angles and timing information has been developed:
the profile constrained fit (PCF).  The PCF uses the one-to-one
correlation between in-plane angle and shower profile: the in-plane
angle with the best fit shower profile is chosen as the in-plane angle
of the shower.  The Gaisser-Hillas function is used in this profile
fit.  The PCF works poorly for events close to the detector (within
about 5 km), and for lower energy events (below $3 \times 10^{18}$
eV), where less of the shower profile is seen.  These events are
excluded from the HiRes-I monocular sample.  The PCF also works poorly
if too much \v{C}erenkov light contaminates the fluorescence signal;
these events are cut also.  In reconstructing MC events, it is found
that, even with these cuts, the resolution is somewhat worse than for
HiRes-II, and that there is an energy bias.

\begin{figure}
  \includegraphics[width=\columnwidth]{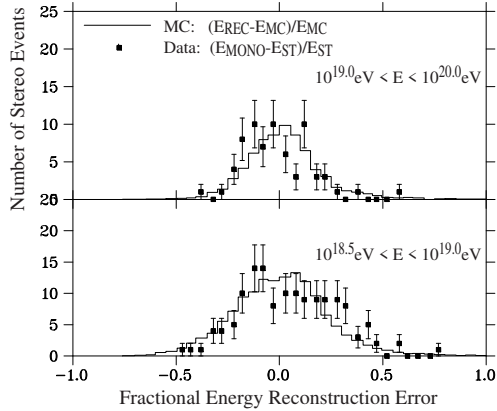}
  \caption{Energy resolution using PCF, after bias correction.  The
    histogram shows MC resolution, the data points show the data
    monocular resolution in stereo events.  For the MC, $E_{REC}$
    refers to the reconstructed, monocular energy, while $E_{MC}$
    refers to the generated energy in the same events.  For the data,
    $E_{MONO}$ refers to the energy reconstructed using the monocular
    geometry (corresponding to $E_{REC}$ in the MC), while $E_{ST}$
    refers to the energy reconstructed using the stereo geometry.}
  \label{fig:hr1_over}
\end{figure}

Since stereo events are seen in both detectors, they have excellent
geometrical determination using the intersection of the two
shower-detector planes.  For these events, comparison of the PCF
reconstruction to the stereo reconstruction shows the same energy
resolution and bias as seen in the MC sample.  Having confidence that
we understand the PCF, we correct for the bias.
Figure~\ref{fig:hr1_over} shows the energy resolution of the PCF
reconstruction for MC events and for stereo events after the
correction.  The agreement is excellent.

\begin{figure}
  \includegraphics[width=\columnwidth]{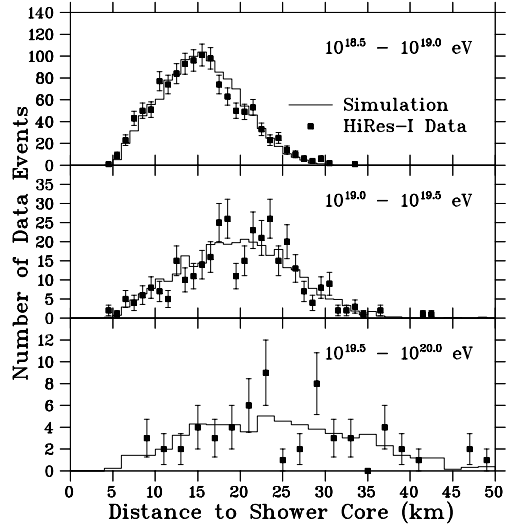}
  \caption{Shower core distance distribution using PCF, in HiRes-I
    data and in MC.  The squares with error bars are the data, the
    histogram is the MC.}
  \label{fig:hr1_dc}
\end{figure}

Figure~\ref{fig:hr1_dc} shows comparisons between the HiRes-I data and
the MC simulation for the distance to the shower core of showers in
three energy bands.  Again the agreement is excellent.

\section{Systematic Uncertainties}

The largest systematic uncertainties in the calculations of energy are
the absolute calibration of the phototubes ($\pm 10\%$)\cite{hr_cal},
the fluorescence yield ($\pm 10\%$)\cite{fl_yield}, and the correction
for unobserved energy in the shower ($\pm
5\%$)\cite{csong,asz_thesis}.  These three uncertainties, added in
quadrature, give an uncertainty in the energy of $\pm15\%$.  This
effect of this energy uncertainty in calculating the flux is
$\pm27\%$\cite{prl}.

To test the sensitivity of the flux measurement to atmospheric
uncertainties, we generated new MC samples with VAOD values of 0.02
and 0.06, i.e., with the average plus and minus one RMS value, and
analyzed them (and the data) using the same VAOD values.  This
provides a conservative estimate of the flux uncertainty since the
systematic uncertainty in the average VAOD is less than the RMS.  The
result was a change in the flux of $\pm 15\%$.  Adding this in
quadrature with the sources of systematic uncertainty described above
results in a net uncertainty of $\pm 31\%$.  This uncertainty is
common to the flux measurements from HiRes-I and HiRes-II.

The effect of using an average VAOD value, rather than the changing
but measured values has also been studied\cite{andreas,asz_thesis}.
The changes to the energy scale and flux are negligible.

The limited elevation coverage of the HiRes-II detector makes the
aperture calculation sensitive to the composition assumptions used in
the MC simulation.  This and other sources of systematic uncertainty
(the given input spectrum and using an average atmosphere) are
considered in reference \cite{andreas}.  The composition assumptions
have a negligible effect on the aperture above an energy of $10^{18}$
eV, and give a systematic uncertainty of order the statistical
uncertainty only at $3\times10^{17}$ eV.

\section{Results}

\begin{figure}
  \includegraphics[width=\columnwidth]{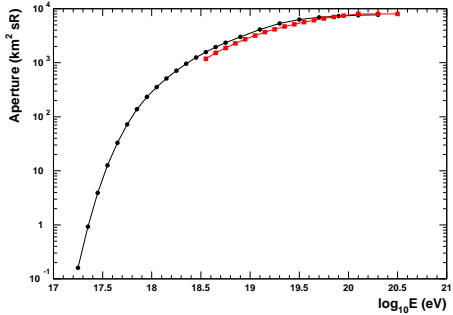}
  \caption{The calculated apertures of HiRes-I (red squares) and
    HiRes-II (black circles) as a function of energy.}
  \label{fig:aper}
\end{figure}

Figure~\ref{fig:aper} shows the calculated aperture of the two HiRes
detectors.  At an energy of $10^{20}$ eV the aperture is nearly 10,000
km$^2$ sr.

\begin{figure*}
  \begin{center}
  \includegraphics[width=1.5\columnwidth]{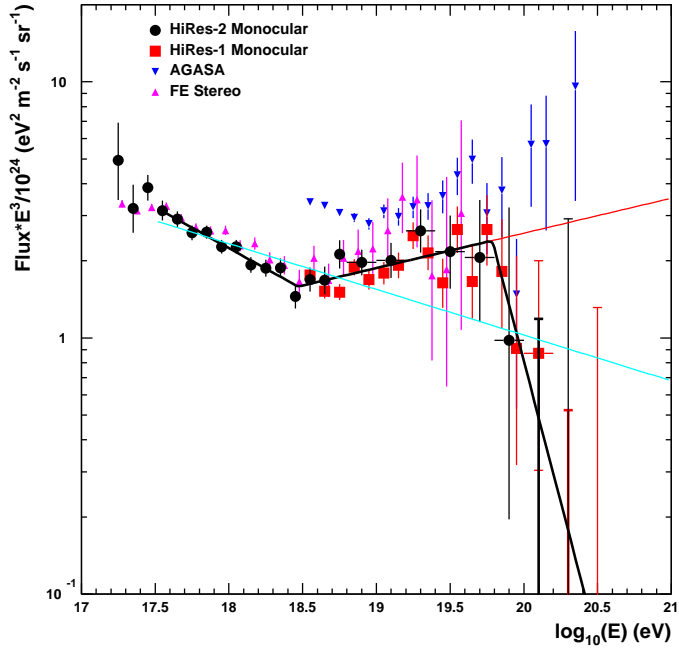}
  \end{center}
  \caption{$E^3$ times the UHECR Flux.  Results from the HiRes-I (red
    squares) and HiRes-II (black circles) detectors, the AGASA
    experiment (blue down-triangles) and the Fly's Eye experiment (in
    stereo mode; magenta up-triangles) are shown.  The Fly's Eye
    points have been shifted to the right by one quarter bin
    ($\Delta\log_{10} E=0.025$) for clarity.  Also shown are two
    spectral law fits to the HiRes-I and HiRes-II spectra as described
    in the text.  The 1$\sigma$ upper limits for two empty bins of
    each HiRes spectra are also shown.}
  \label{fig:je3}
\end{figure*}

Figure~\ref{fig:je3} shows the measured spectrum of cosmic
rays\cite{webtable}.  The spectrum has been multiplied by E$^3$ for
clarity.  The closed squares (open circles) are the HiRes-I (HiRes-II)
measurements.  For comparison to previous experiments, the
up-triangles are the stereo Fly's Eye spectrum\cite{fester}, and the
down-triangles are the result of the Akeno Giant Air Shower Array
(AGASA)\cite{agasa}.  The HiRes-I and HiRes-II monocular measurements
agree with each other very well in the overlap region, and are also in
good agreement with the Fly's Eye stereo spectrum.

In this plot the ankle shows up clearly at $3\times10^{18}$ eV
($\log_{10}E=18.5)$.  The spectrum steepens again at $6\times10^{19}$
eV ($\log_{10}E=19.8)$.  The AGASA spectrum appears to continue
unabated above this energy while the HiRes spectrum falls above this
point.

We test whether our data are consistent with this interpretation of
the AGASA spectrum by fitting our data to a broken power law.  This
fit is also shown on Figure~\ref{fig:je3}.  This fit had two floating
break points separating three regions of constant spectral slope.  The
fit was performed using the normalized, binned maximum likelihood
method\cite{cousins}, which allows us to include sparsely populated
and empty bins.  The fitted break points are at $\log_{10}E\;\text{(in
  eV)} = 18.47\pm0.06$ and $19.79\pm0.09$.  The fitted spectral slopes
are $\gamma = 3.32\pm0.04$, $2.86\pm0.04$, and $5.2\pm1.3$.  The
$\chi^2$ for the fit is 30.1 for 33 degrees of freedom.  If we extend
the middle section of the fit (as shown by the red/grey line in
Figure~\ref{fig:je3}) to higher energies, our aperture predicts that
we should have 28.0 events above the second break point energy of
$\log_{10}E = 19.79$, where we really have 11.  The Poisson
probability for 11 or fewer events with a mean of 28 is $2.4 \times
10^{-4}$.  We therefore conclude that our data is not consistent with
a continuation of the spectrum unabated above the pion production
threshold.  It is worth emphasizing that we have considerable
sensitivity to such a continuation, but the data do not support it.

A similar fit with only one break point has a $\chi^2$ of 46.0 for 35
degrees of freedom, worse by nearly 16 than the fit above (a
significance of $\sim3.7\sigma$).  The fitted break point is at
$\log_{10}E = 18.45\pm0.03$, and the fitted spectral slopes are
$\gamma = 3.32\pm0.03$ and $2.85\pm0.05$.  A fit with no break points
at all has a bad $\chi^2$ of 114 for 37 degrees of freedom,
demonstrating that the ankle is clearly observed in our data.  The
spectral slope is $\gamma = 3.12\pm0.01$.  This fit is shown on
Figure~\ref{fig:je3} as a cyan/light grey line.

\section{Fitting the Spectrum}

The implications of our spectrum measurement can be explored using a
toy model of UHECR.  In this model, there are two types of sources,
galactic and extragalactic.  We choose the galactic sources to be
consistent with the HiRes/MIA and HiRes stereo composition
measurements\cite{hiresmia,gregpaper}: we assign the iron component of
the cosmic ray flux to be galactic\cite{waxman}.  This assignment is
consistent with the expectation that the highest energy galactic
cosmic rays should be those of the highest charge.  The proton
component we take to be extragalactic.

\begin{figure}
  \includegraphics[width=\columnwidth]{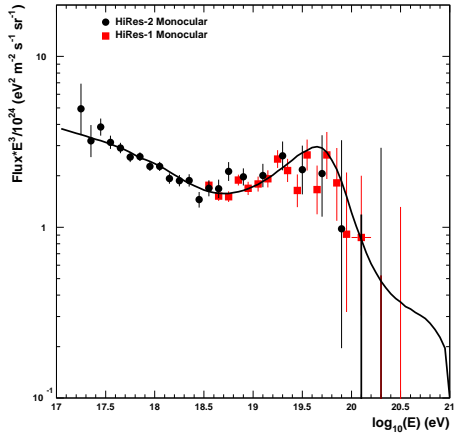}
  \caption{$E^3$ times the UHECR Flux.  Results from the HiRes-I (red
    squares) and HiRes-II (black circles) detectors are shown.  Also
    shown is a fit to a model described in the text.  The 1$\sigma$
    upper limits for two empty bins of each HiRes spectra are also
    shown.}
  \label{fig:je3fit}
\end{figure}

To describe the extragalactic cosmic rays, we assume that all sources
have the same power law spectrum, and that cosmic rays lose energy in
propagating to the earth by pion and $e^+ e^-$ production from the
CMBR photons, and by the cosmological red shift\cite{berezinsky}.  The
sources are assumed to be uniformly distributed and to evolve in
density by $(1+z)^m$.  Figure~\ref{fig:je3fit} shows our spectrum
result with the best fit superimposed on it.  The fitted values $m$
and of $-\gamma$, the spectral slope of the spectrum at the source,
are $m = 2.6\pm0.4$ and $-\gamma = 2.38\pm0.05$.

\section{Summary}

We have measured the flux of ultrahigh energy cosmic rays from
$1.6\times10^{17}$ eV to over $10^{20}$ eV.  Our experiment detects
atmospheric fluorescence light from cosmic ray showers and performs a
calorimetric measurement of cosmic ray energies.  We perform
calibrations of our detector and measure the light-scattering
properties of the atmosphere.  The total systematic uncertainty in our
spectrum measurement averages 31\%.

In our energy range we observe two features in the UHECR spectrum
visible through changes in the spectral power law.  We observe the
ankle at $3\times10^{18}$ eV.  We also have evidence for a suppression
at a higher energies, above $6\times10^{19}$ eV.

This work is supported by US NSF grants PHY-9321949, PHY-9322298,
PHY-0098826, PHY-0245428, PHY-0305516, PHY-0307098, by the DOE grant
FG03-92ER40732, and by the Australian Research Council.  We gratefully
acknowledge the contributions from the technical staffs of our home
institutions and the Utah Center for High Performance Computing.  The
cooperation of Colonels E. Fischer and G. Harter, the US Army, and the
Dugway Proving Ground staff is greatly appreciated.

\end{document}